\documentstyle[psfig]{mn}

\newif\ifAMStwofonts

\def\mum{\mu\rm{m}}
\def\deg{^\circ}
\def\min{^{\prime}}
\def\sec{^{\prime\prime}}
\def\jyb{Jy beam$^{-1}$}
\def\jybs{Jy beam$^{-1}$~}

\def\rasec {\hbox{$\,$\raise 0.6 ex \hbox{\rm s}\kern-.35em
                  \lower 0.0 ex \hbox{.}$\,$}}        
\def\decsec{\hbox{$\,$\raise 0.0 ex \hbox{$\sec$}\kern-.45em
                  \lower 0.0 ex \hbox{.}$\,$}}         
\def\decmin{\hbox{$\,$\raise 0.0 ex \hbox{$\min$}\kern-.45em
                  \lower 0.0 ex \hbox{.}$\,$}}  
\def\gtabouteq{\,\hbox{\raise 0.5 ex \hbox{$>$}\kern-.77em 
                    \lower 0.5 ex \hbox{$\sim$}$\,$}}       
\def\ltabouteq{\,\hbox{\raise 0.5 ex \hbox{$<$}\kern-.77em 
                     \lower 0.5 ex \hbox{$\sim$}$\,$}}

\ifoldfss
  \ifCUPmtlplainloaded \else
    \NewTextAlphabet{textbfit} {cmbxti10} {}
    \NewTextAlphabet{textbfss} {cmssbx10} {}
    \NewMathAlphabet{mathbfit} {cmbxti10} {} 
    \NewMathAlphabet{mathbfss} {cmssbx10} {} 
  \fi
  \ifAMStwofonts
    \ifCUPmtlplainloaded \else
      \NewSymbolFont{upmath} {eurm10}
      \NewSymbolFont{AMSa} {msam10}
      \NewMathSymbol{\upi}     {0}{upmath}{19}
      \NewMathSymbol{\umu}     {0}{upmath}{16}
      \NewMathSymbol{\upartial}{0}{upmath}{40}
      \NewMathSymbol{\leqslant}{3}{AMSa}{36}
      \NewMathSymbol{\geqslant}{3}{AMSa}{3E}

      \let\leq=\leqslant 
       
    \fi
  \fi
\fi 

\ifnfssone
  \newmathalphabet{\mathit}
  \addtoversion{normal}{\mathit}{cmr}{m}{it}
  \addtoversion{bold}{\mathit}{cmr}{bx}{it}
  \newmathalphabet{\mathbfit} 
  \addtoversion{normal}{\mathbfit}{cmr}{bx}{it}
  \addtoversion{bold}{\mathbfit}{cmr}{bx}{it}
  \newmathalphabet{\mathbfss} 
  \addtoversion{normal}{\mathbfss}{cmss}{bx}{n}
  \addtoversion{bold}{\mathbfss}{cmss}{bx}{n}
  \ifAMStwofonts
    \ifCUPmtlplainloaded \else
      \UseAMStwoboldmath
      \makeatletter
      \new@mathgroup\upmath@group
      \define@mathgroup\mv@normal\upmath@group{eur}{m}{n}
      \define@mathgroup\mv@bold\upmath@group{eur}{b}{n}
      \edef\UPM{\hexnumber\upmath@group}
      \new@mathgroup\amsa@group
      \define@mathgroup\mv@normal\amsa@group{msa}{m}{n}
      \define@mathgroup\mv@bold\amsa@group{msa}{m}{n}
      \edef\AMSa{\hexnumber\amsa@group}
      \makeatother
      \mathchardef\upi="0\UPM19
      \mathchardef\umu="0\UPM16
      \mathchardef\upartial="0\UPM40
      \mathchardef\leqslant="3\AMSa36
      \mathchardef\geqslant="3\AMSa3E

      \let\leq=\leqslant 

    \fi
  \fi
\fi 

\ifnfsstwo
  \DeclareMathAlphabet{\mathbfit}{OT1}{cmr}{bx}{it}
  \SetMathAlphabet\mathbfit{bold}{OT1}{cmr}{bx}{it}
  \DeclareMathAlphabet{\mathbfss}{OT1}{cmss}{bx}{n}
  \SetMathAlphabet\mathbfss{bold}{OT1}{cmss}{bx}{n}
  \ifAMStwofonts
    \ifCUPmtlplainloaded \else
      \DeclareSymbolFont{UPM}{U}{eur}{m}{n}
      \SetSymbolFont{UPM}{bold}{U}{eur}{b}{n}
      \DeclareSymbolFont{AMSa}{U}{msa}{m}{n}
      \DeclareMathSymbol{\upi}{0}{UPM}{"19}
      \DeclareMathSymbol{\umu}{0}{UPM}{"16}
      \DeclareMathSymbol{\upartial}{0}{UPM}{"40}
      \DeclareMathSymbol{\leqslant}{3}{AMSa}{"36}
      \DeclareMathSymbol{\geqslant}{3}{AMSa}{"3E}

      \let\leq=\leqslant 

    \fi
  \fi
\fi 

\ifCUPmtlplainloaded \else
  \ifAMStwofonts \else 
    \def\upi{\pi}
    \def\umu{\mu}
    \def\upartial{\partial}
  \fi
\fi

\title{SCUBA Observations of NGC~1275}
\author[Judith A.Irwin et al.]
      {Judith A. Irwin,$^1$
       J. M. Stil,$^1$
       and T. J. Bridges$^2$\\
       $1$Dept. of Physics, Queen's University, Kingston, Canada, K7L 3N6\\
       $2$Anglo-Australian Observatory, PO Box 296, Epping, NSW 2121, Australia}
\date{Accepted xxx.
      Received xxx;
}

\pagerange{\pageref{firstpage}--\pageref{lastpage}}
\pubyear{2000}

\begin{document}

\maketitle

\label{firstpage}

\begin{abstract}

Deep SCUBA observations of NGC~1275 at 450$\mum$ and 850$\mum$ along
with the application of deconvolution algorithms have permitted  us to
separate the strong core emission in this galaxy from the fainter extended
emission around it.  The core has a steep spectral index
and is likely due primarily to the AGN.  
The faint emission has a positive spectral index and
is clearly due to extended dust in a patchy distribution out to a radius
of $\sim$ 20 kpc from the nucleus.  These observations have now revealed
that a
large quantity of dust, $\sim$ 6 $\times$ 10$^7$ $M_\odot$, 2 orders of
magnitude larger than that inferred from
previous optical absorption measurements,
 exists in this galaxy.  We estimate the temperature of this dust to
be $\sim$ 20 K (using an emissivity index of $\beta$ = 1.3) and the
 gas/dust ratio to be 360.  These values are typical of
spiral galaxies.
The dust emission correlates spatially with the hot X$-$ray emitting gas
which may be due to
collisional heating of broadly distributed dust
by electrons. 
Since the destruction timescale
is short, 
the dust cannot be replenished by stellar mass loss and must be
externally supplied, either via the infalling galaxy
or the cooling flow itself.

\end{abstract}


\section{Introduction}
\label{intro}

Until recently, it was not thought that centrally-located cluster galaxies
contained significant amounts of dust.  However, recent observations have
begun to reveal large amounts of dust in some central galaxies.  The
existence of dust is inferred from a diverse set of observations,
including the ratios of emission lines in optical filaments (Hu 1992; Donahue
\& Voit 1993; Crawford et al. 1999), dust lanes in optical images (e.g.
McNamara, O'Connell \& Sarazin 1996; Pinkney et al. 1996), near infrared (NIR)
 imaging and
spectroscopy (e.g. Krabbe et al. 2000; Donahue et al. 2000; Jaffe, Bremer, \&
van der Werf
2000), and X$-$ray absorption (e.g. White et al. 1991; Allen \& Fabian 1997;
Arnaud \& Mushotzsky 1998; Allen 2000).

Searches at longer wavelengths have now also 
revealed emission from dust directly, especially cooler dust 
which may be much more plentiful than the warmer components. 
The Infrared Astronomical Satellite
(IRAS) detected several central cluster galaxies at 60 and 100
$\mu$m; Cox, Bregman, \& Schombert (1995) find that $\sim$ 10\% of central galaxies have
far infrared (FIR)
emission, with dust masses of $\sim$ 10$^7$ M$_{\odot}$.  More
recently, Lester et al. (1995) have detected 100 $\mu$m emission from NGC
1275 with the Kuiper Airborne Observatory (KAO), 
and there have been Infrared Space Observatory (ISO)
detections at 60$-$200 $\mu$m
of a few central galaxies (e.g. Hansen et al. 2000).  Annis \& Jewitt
(1993) carried out an unsuccessful search for dust emission in 11 cooling
flow clusters using the UKT-14 bolometer system on the 
James Clerk Maxwell Telescope (JCMT).  More
recently, however, Edge et al. (1999) have detected dust emission at 850 $\mu$m in
two cooling flow clusters using the more sensitive Submillimetre 
Common-User Bolometer Array (SCUBA) camera.

The origin of the dust in central cluster galaxies is a matter of
some controversy.  Central cluster galaxies live in complicated
environments, with many at the centers of cluster cooling flows
(e.g. Fabian 1994), and interactions/mergers may also be common.  
Dust in central cluster galaxies may thus originate in several ways: from
the cooling flow itself or in cold, dense clouds condensed from the flow;
from ongoing star formation, perhaps initiated by the cooling flow; or from the
merger of a gas/dust rich galaxy.  Direct detections of dust in central
cluster galaxies, and good determinations of the dust properties (mass,
temperature, and correlations with other components) will help enormously
to decide between these alternatives.  For example, the bulk of the dust
in cooled gas clouds is expected to be quite cold with T $<$ 15K (e.g.
Fabian, Johnstone, \& Daines 1994; 
Johnstone, Fabian \& Taylor 1998) [However, this conclusion is
disputed by O'Dea, Baum \& Gallimore (1994) and Voit \& Donahue (1995), who claim
minimum dust temperatures of $\sim$ 20 K.]

The very existence of dust in central cluster galaxies is in itself
surprising, since any dust should be destroyed quickly by sputtering from
X$-$rays from the hot cluster gas, on timescales $\leq$ 10$^7$ years (e.g.
Draine \& Salpeter 1979).  Either the dust is shielded against sputtering
(e.g. at the centers of cold, dense molecular clouds), or the dust is
replenished on similar timescales.  This issue is discussed in more detail
in Sect.~\ref{origin}.

NGC 1275 is a complex and fascinating object.
It  has been studied in
virtually every waveband, including the optical (e.g.
N{\o}rgaard-Nielsen et al.
1993; McNamara et al. 1996), NIR (e.g. Donahue et al. 2000;
Krabbe et al. 2000) and far infrared (FIR) (Gear et al. 1985;
Lester et al. 1995).
It has been classified as a Seyfert (Seyfert
1943)
and has an active galactic nucleus (AGN) in the form of
the highly variable radio source 3C 84 (e.g. Pedlar et
al. 1990).  Minkowski (1957) discovered two distinct filamentary systems
around NGC 1275.  The low velocity system (LVS) has the same velocity
($\sim$ 5300 km s$^{-1}$) as that of NGC 1275 itself, and has been studied in
detail by several authors (e.g. Lynds 1970; Hu et al. 1983; Caulet et al.
1992).  The high velocity system (HVS) has a velocity $\sim$ 3000 km s$^{-1}$
higher than NGC 1275 and consists of giant HII regions to the North and
North-West of the galaxy center; it has generally been concluded that the
HVS is a late-type spiral galaxy.  At least part of the HVS is in front of
the nucleus, as determined
from HI absorption against the nuclear radio source (De Young, Roberts, \&
Saslaw
 1973) and Ly$\alpha$ absorption (Briggs,
Snijders, \& Boksenberg 1982).  The HVS
could be merely a chance projection onto NGC 1275, but several authors
have argued that the HVS is in fact colliding with NGC 1275 (e.g.  Hu et
al. 1983; Unger et al. 1990; Caulet et al. 1992; N{\o}rgaard-Nielsen et al.
1993).  Recent observations of gas at velocities intermediate between the
LVS and HVS also strongly argue for a physical connection between
the two systems (Ferruit et al. 1998).

  It has been known for some time that the Perseus
cluster hosts a large cooling flow (e.g. Fabian et al. 1981); analysis of
recent 
Advanced Satellite for Cosmology and Astrophysics (ASCA) data
show that the mass infall rate is $\sim$ 300
M$_{\odot}$ yr$^{-1}$ (Allen et al. 1999). R\"ontgen Satellite (ROSAT)
 and Chandra X$-$ray observations
of NGC 1275 and the Perseus cluster (B{\"o}hringer et al. 1993, and Fabian et
al. 2000, respectively) have also revealed two holes in the NGC 1275 X$-$ray
emission, thought to be caused by displacement of hot gas by the radio
lobes.  
NGC 1275 is the only cooling flow central galaxy
with detected CO emission, with an inferred molecular H$_2$
mass of
$\sim$ 10$^{10}$M$_{\odot}$; see Bridges \& Irwin (1998) for a summary of
the many previous CO studies.

In this paper, we present new SCUBA 450/850 $\mu$m data for
NGC 1275, in an attempt to study the cold dust in this galaxy in more
detail.  We are especially interested in such dust properties as:  its
spatial distribution and correlation with other components, total mass,
temperature and 
ultimately, its origin.  In Sect.~\ref{obs_data}, we present the observations
and data reductions, unprocessed and processed images, Sect.~\ref{results}
presents the results, including an analysis of the AGN and extended dust 
components, separately, and correlations with other wavebands,
 Sect.~\ref{discussion} discusses the origin of
the dust and dust heating, and Sect.~\ref{conclusions} presents the conclusions.
For NGC~1275, we take D = 70 Mpc (H$_0$ = 75 km s$^{-1}$ Mpc$^{-1}$, 
Bridges \& Irwin 1998). 

\section[]{Observations and Data Reduction}
\label{obs_data}

\subsection{Maps of NGC~1275}
\label{maps}


Table~\ref{obslog-tab} lists the observing log of SCUBA jiggle maps
obtained on July 15, October 31 and December 14 1998. The total
observation time was 5.5 hours, half of which was spent on the source.
Chopping was performed in Azimuth, with a chop thow of 120 arcseconds.

The standard reduction scheme for SCUBA jiggle maps was followed, consisting
of flatfielding, correction for atmospheric extinction, removal of
noisy bolometers, sky noise removal and despiking. These steps are
described in detail in Sandell (1997). Flatfielding
corrects for differences in sensitivity of the bolometers.  The
atmospheric opacity at 450 and 850 $\mum$ was determined from skydip
observations. 
Noisy bolometers were identified from noise measurements
performed in between observations, and visual inspection of the signal
of all bolometers as a function of time during the observations.

Variations of the sky brightness are not canceled out completely
by chopping because the atmosphere in the direction of the chopped beam
is significantly different from the atmosphere in the direction of the 
source.  This is a particular
concern for observations of an extended source, where a large chop
throw is required (Holland et~al. 1999).  Variations of the sky on
short time scales were further eliminated by subtracting the median
sky level measured by off-source bolometers for each 18 second
integration. Although the sky level is determined with higher accuracy
if more bolometers are considered, the likelihood that faint extended
emission of the source is also subtracted, increases as well. A
compromise must be sought between decreasing the sky noise and the
possibility of subtracting source emission.  The sky level was
determined from the outer two rings of bolometers at 450 $\mum$, and
the single outer ring of bolometers at 850 $\mum$.  In this way a
significant aperture around the centre of the field of view is
excluded from the determination of the sky level. The radius of the
aperture which is never covered by any of the ``sky''-bolometers is
$28''$ at 450 $\mum$ and $46''$ at 850 $\mum$. As the sky level at any
time is taken to be the median of a ring of 54 (450 $\mum$) or 18 (850
$\mum$) bolometers surrounding the source, only a smooth symmetric
component would be subtracted in this way. 

Occasional spikes in the output of each bolometer
were then removed by excluding signals deviating more than $\pm 3 \sigma$ from 
the mean signal.

\begin{figure*}
\psfig{figure=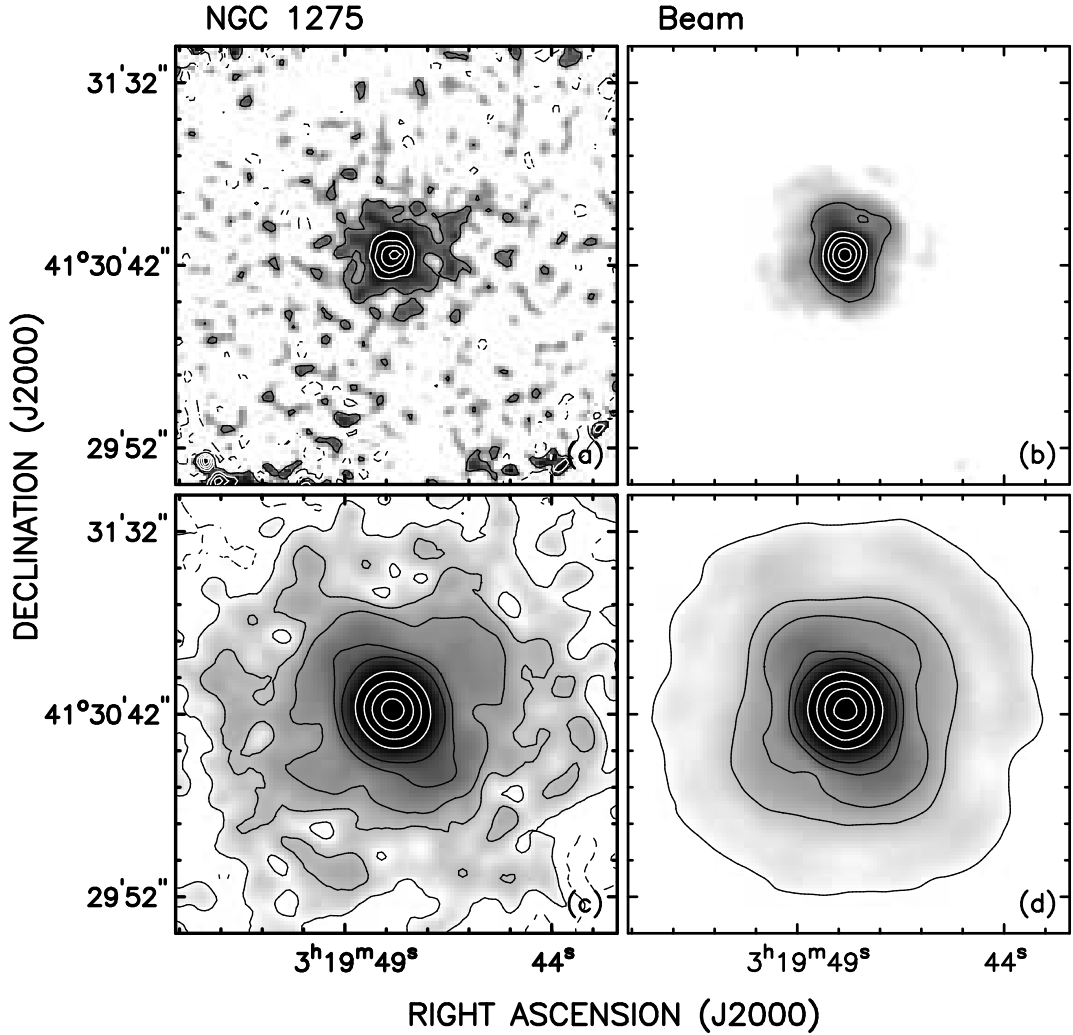,angle=0,width=\textwidth}
\caption{Maps of NGC~1275 and the coadded beam maps at 450$\mum$ (top row)
and 850$\mum$ (bottom row). Contours in
the 450$\mum$ map of NGC~1275 are drawn at $-10$, $10$, $30$, $50$,
$70$, and $90$\% of the peak intensity (0.78 Jy/beam). Grayscales
start at 3\% level on a logarithmic scale. Contours in the 850$\mum$
map are drawn at $-0.3$, $0.3$, $1$, $2$, $5$, $10$, $30$, $50$, $70$, $90$\%
of the peak intensity (1.44 Jy/beam).  Grayscales start at $0.3$\% of
the peak intensity. The contours and grayscales in the beam maps
represent the same percentiles of the peak intensity as in the
corresponding maps of NGC~1275. 
\label{maps-fig}
}
\end{figure*}

The data were averaged with weight factors $w_i= (t_i/N_i)^2$ 
where $t_i$ and $N_i$ are the integration time
and noise level of observation $i$,
to
produce the maps shown in Fig.~\ref{maps-fig}. These weight factors
take into account both the different integration times and
atmospheric opacities of the observations, since the noise in each
map is dominated by the sky. The relative weights of the three nights
are July:October:December=0.48:1:0.64.

Recently, a potential calibration problem resulting from erroneous
values of the hot and cold load temperatures was identified (see
{\texttt{http://www.jach.hawaii.edu/JACdocs/JCMT/SCD/SN/002/\break
tau$\_$analysis.html}}). 
This problem could affect the sky opacity
derived from skydip observations. The calibration was repeated with
opacities calculated from the CSO opacity and the revised relations
for the opacities at $450\mum$ and 850$\mum$. We found that 
this procedure had no
effect on our calibration.

Taking into account the repeatability of the flux determination over
a variety of integration times as well as an assessment of the accuracy of
the calibration factor
including uncertainties in the calibrator fluxes, 
we estimate that the 450$\mum$ calibration
error is $\sim$ 25\% and the 850$\mum$ error is $\sim$ 10\%.  Typical
pointing corrections were 1 to 2 arcseconds over the course of an hour
and residual pointing errors will be smaller than this.

The corrected, calibrated maps are shown in Fig.~\ref{maps-fig}a (450$\mum$)
and Fig.~\ref{maps-fig}c (850$\mum$).

\subsection{Beam maps}

Beam maps were constructed for each night from calibrator maps with
the same chop throw that was applied to the observations of NGC~1275.
These calibrator maps were reduced in the same way as the observations
of NGC 1275, with one exception: noisy bolometers were not excluded
from the analysis. 
This was done to avoid the occurrence of unsampled areas in the
planet maps.
 However no significant noise was introduced, because at
850$\mum$ the primary calibrators Mars and Uranus are brighter than
NGC~1275 by factors of 300 and 60 respectively.  At 450$\mum$ these
factors are even larger because the brightness of the planets rises,
whereas the spectrum of NGC~1275 shows a decline with frequency. 
 Therefore,
it is safe to conclude that noisy bolometers in the beam maps do not
affect our results.  

Calibrator sources are Uranus, Mars, CRL~618, and HL~Tau. We used
planet maps made with the same chop throw as the NGC~1275 maps, to
define the shape of the beam for each night.  Only for the 450$\mum$
map in December, we used the secondary calibrator CRL~618, as no
450$\mum$ planet map with the appropriate chop throw was available.
The calibration factors adopted here are those determined for the
October observations: 623.69 Jy/(Volt beam) at 450$\mum$ and
251.54 Jy/(Volt beam) at 850 $\mum$.  Our largest calibrator was Mars
 which was smaller than
5$\sec$ over the course of the observations.  This broadens the beam
by less than 0.3$\arcsec$ at 850$\mum$ and by less than 0.6$\arcsec$
at 450$\mum$, or $<$ 1.8\% and $<$ 6.7\% of the beam size, respectively
(see caption to Figure 2 for the fitted beam size).  Thus, 
the finite size of the calibrators will not significantly
affect the beam, or its far sidelobes.

The coordinate system for the beam maps produced in this way was
azimuth-elevation because chopping was done in azimuth, and the beam 
tends to be 
slightly elongated in the direction of chopping.
For each observation of NGC~1275, the appropriate
beam map was shifted to the position of the peak in the final map of
NGC~1275, and rotated with the field rotation angle at the time of
mid-observation.  The resulting beam maps were coadded with the same
weight factors as the data for NGC~1275 
(see Sect.~\ref{maps}) to construct a
map of the effective beam.  The beam
maps are shown in Fig.~\ref{maps-fig}b and d.  The most distinctive
elongated feature in position angle $\sim 40$ degrees visible in the
850$\mum$ maps of NGC~1275 and the final beam map, can be identified
with the chop direction of the October 31 observation, which received
a relatively high weight due to the favourable atmospheric conditions.

The resulting beam maps (Fig.~\ref{maps-fig}b and d)
give the best possible representation of the JCMT beam
applicable to the coadded maps of NGC~1275. As only one
calibrator observation with the correct chop throw was made in a
night, we cannot correct for variations of the error lobes during the
night. The main source of such variations is the sensitivity of the
structure of the telescope to temperature variations. Large
changes in the structure of the error lobes have been reported in the
early evening (until 9pm) and in late morning (after 8am). 
 However, only one observation, the 850$\mum$ beam
map of December 14, was made outside the relatively stable window
between 9pm and 8am. 

\begin{table}
\begin{tabular}{|  l  | l  | c  | c  | c  | c  |} 
\hline 
Num  &  Date    &  Time     &  Exp.time. &  Airmass  &  $\tau_{850}$ \\ 
     &          &  (Local)  &    (hour)    &         &          \\ 
\hline
130  &  Jul 15  &   05:25   &    0.5     &  1.276    &   0.23   \\ 
131  &  Jul 15  &   05:53   &    0.5     &  1.202    &   0.23   \\ 
134  &  Jul 15  &   06:29   &    0.5     &  1.135    &   0.26   \\ 
135  &  Jul 15  &   06:58   &    0.5     &  1.102    &   0.26   \\ 
144  &  Jul 15  &   08:08   &    0.5     &  1.079    &   0.29   \\ 
145  &  Jul 15  &   08:36   &    0.5     &  1.091    &   0.29   \\ 
148  &  Jul 15  &   09:12   &    0.5     &  1.128    &   0.30   \\ 
12   &  Oct 31  &   02:11   &    1       &  1.160    &   0.16   \\ 
57   &  Dec 14  &   00:23   &    1       &  1.326    &   0.18   \\ 
\hline 
\end{tabular} 
\caption{Observation log for SCUBA 450/850 $\mum$ observations of NGC~1275.
\label{obslog-tab}}
\end{table}

\subsection{Processed Images}

\begin{figure*}
\psfig{figure=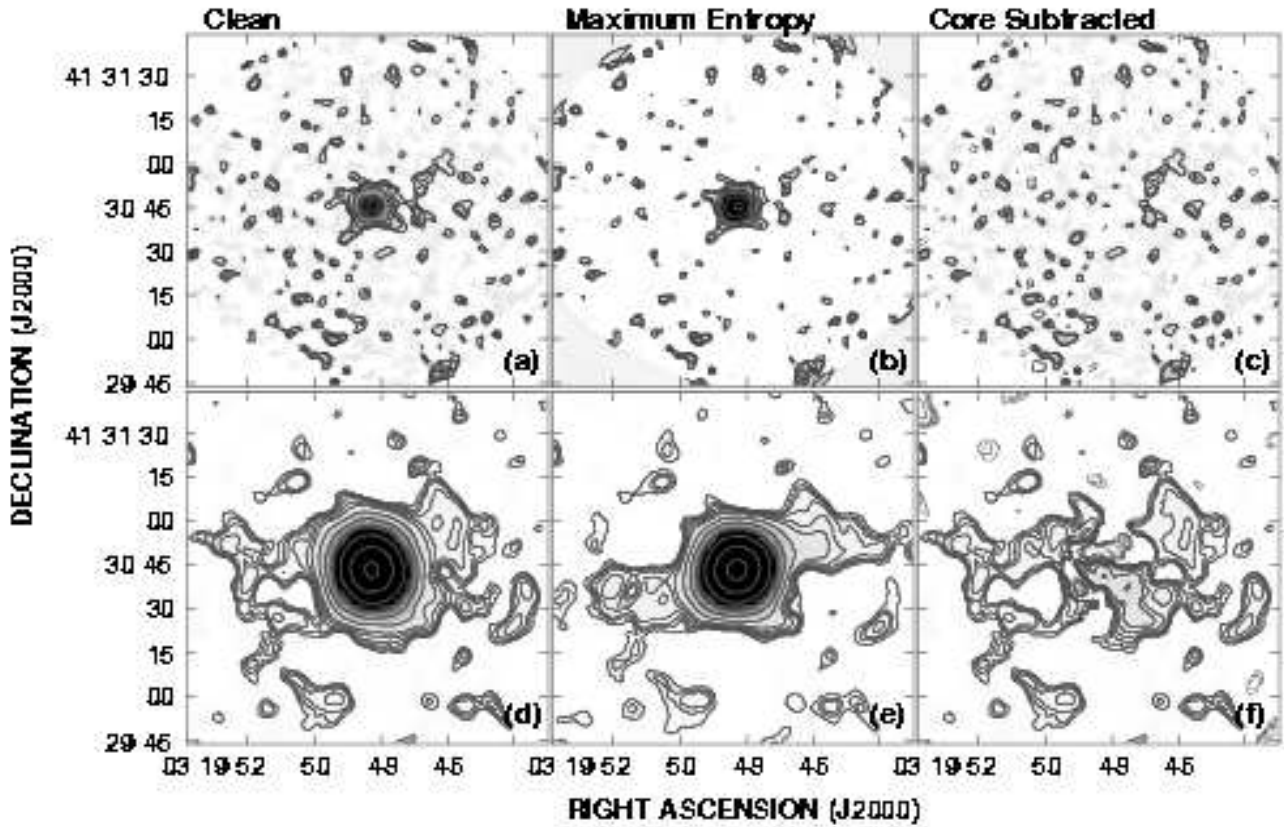,width=\textwidth}
\caption{Processed (clean, maximum 
 entropy and core-subtracted) maps of NGC~1275.{\bf(a)}  Clean map at 450$\mum$.
The image was cleaned to the first negative component and all cleaned components
are associated with the strong central source.  The greyscale ranges from 0 to 0.7 
\jybs and contours are at 75 (2$\sigma$), 100, 140, 225, 450, and 700 m\jyb
with a peak at 756 m\jyb.  The clean beam size is
9.1$\sec$ $\times$ 8.8$\sec$ @ position angle PA = 63.3$\deg$.
{\bf (b)}  Maximum entropy map at 450 $\mum$.  
Contours are at 28, 50, 100, 200, 400, and 650
m\jybs and the greyscale is from 0 to 0.35 \jyb.  The beam is the same as in (a).
{\bf (c)}  Residual map formed from map (a) but with the strong central source subtracted.
The greyscale ranges from 0 to 0.7 \jybs and contours are at
-100, -75, 75 and 100 m\jyb.  
{\bf (d)} Clean map at 850$\mum$ using the same procedure as at 450$\mum$.  The greyscale
ranges from 0 to 0.3 Jy beam$^{-1}$ and the contours are at 
7 ($\approx$ 2$\sigma$), 8.5, 10, 13, 18, 25, 50, 100, 200, 400, 800, and 1350 mJy beam$^{-1}$.
The peak flux is 1.44 Jy beam$^{-1}$.  The clean beam has 
parameters, 17.1$\sec$ $\times$ 15.7$\sec$ at PA=1.7$\deg$.
{\bf (e)} Maximum entropy map.  The contours are
the same as in (d) except for the lowest level which has been
set to 6 m\jyb. The  
 greyscale ranges from 0 to 0.32 \jyb.  The beam is the same as in (d).
{\bf (f)}  Residual map after subtracting the central source.  The greyscale ranges from
0 to 0.3 \jybs and contours are at -25, -18, -13, -10, -8.5, -7, 7, 8.5, 10, 13, 18,
 25, and  50 m\jyb.  
}
\label{final-maps-fig}
\end{figure*}

The beam shape begins to show irregular, non-gaussian patterns below 
the 5\% level and it is desirable to remove such features from the total intensity
image in order to see whether real residual emission remains.  We attempted to
do this by subtracting the scaled beam from the image; however, this assumes that
the strong central source due to the AGN is strictly a point source and the result
leaves some significant residual emission at the map center.  We therefore
treated the image like a typical radio map and applied deconvolution methods
to remove the ``dirty beam" and replace it with a ``clean beam" with the same
characteristics.  The resulting maps retain the information from both the strong
central source as well as any fainter extended emission, while removing the
low level beam irregularities.  
Results using the CLEAN algorithm (Clark 1980), using the Astronomical Image Processing
System (AIPS) routine, APCLN,  are shown
in Fig.~\ref{final-maps-fig}a 
(450$\mum$) and Fig.~\ref{final-maps-fig}d (850$\mum$) and results 
using the maximum entropy (ME) method (Cornwell \& Evans 1985) 
using the AIPS routine, VTESS, are shown in 
Fig.~\ref{final-maps-fig}b and Fig.~\ref{final-maps-fig}e. 
A comparison of these maps provides an indication as to what emission is real.
As the ME maps are well known to enhance faint, broad-scale structure but are
less accurate than the clean maps in characterizing the emission quantitatively,
we use the clean maps for all subsequent quantitative work.  

Residual
maps, i.e. after subtraction of all clean components, are shown in
Fig.~\ref{final-maps-fig}c and f.  
For these images, we show both negative and positive contours so that
the noise is well illustrated.
 Since all subtracted
clean components are associated with the central bright core, the residual
maps represent core-subtracted emission.
At 450 $\mum$,
the core-subtracted map is dominated by noise which is about 10 times higher
than that of the 850$\mum$ map (see figure caption).  At 850$\mum$ 
(Fig.~\ref{final-maps-fig}f),
the noise is much lower, so some additional
errors associated with subtracting
the central source also show up, the largest of which are within the central
30$\sec$ region.  Since positive residual emission 
 could be real, we take the
negative residual emission 
in this region
to represent the errors from the subtraction
 process.  Applying a 2$\sigma$ cutoff
to the core-subtracted and total intensity maps, we formed a ratio map to
indicate the relative
error in this region, finding an average error of 3.0\%
and a maximum error of 6.7\%.  

Is the resulting emission due to continuum alone?
Bridges \& Irwin (1998) have detected CO(J=3-2) emission in NGC~1275 at points
within  $\pm$ 7$\sec$ of the core. Their CO(J=3-2) beam of 14$\sec$ is similar to our
SCUBA 850$\mum$ beam of 16$\arcsec$.
 The CO line occurs at a redshifted
frequency of 339.9 GHz and is therefore within the 30 GHz-wide SCUBA band which is
centered at 347 GHz. 
Bridges \& Irwin  find an integrated
intensity of 12 K km s$^{-1}$  which corresponds 
 to 8.6 m\jybs  when diluted by the
30 GHz bandwidth (assuming constant response across the band).  The value will be
smaller when the bandpass response is considered (see Holland et al. 1999).  This  
corresponds to
only 0.6\% of the SCUBA flux of 1.42 Jy within
 a 16$\sec$ beam and is well within the noise.  Thus, there is no contamination
by the CO line at the core and we do not expect any contamination at 
locations away from
the core, either, where no CO emission has been detected.

\section{Results}
\label{results}

The processed images (Fig.~\ref{final-maps-fig}) show that the 450 $\mum$ and 850 $\mum$
emission is dominated by the bright AGN in NGC~1275.  However, fainter extended 
emission is
also observed and appears to be real. 
Since the core is quite prominent in the processed maps, it has been
possible to separate it from the extended emission and results from this process, along
with total fluxes and the spectral index, are given in
 Table~\ref{flux-tab}.  

\begin{table*}
\centering
\begin{minipage}{140mm}
\caption{NGC~1275 Measured Parameters}
\label{flux-tab}
\begin{tabular}{@{}lcc@{}}
\hline
Parameter                       &    450$\mum$               &    850$\mum$            \\
\hline
Total Flux (Jy)\footnote{The error in the flux reflects the
calibration error noted in Sect.~\ref{maps}. 
Errors associated with varying the box size to
include or exlude low S/N peaks at larger radii are smaller than this.
}      &  1.2 $\pm$ 0.3  & 1.60 $\pm$ 0.16 \\
Core Flux (Jy)\footnote{
The core parameters were derived from gaussian fits to the core component alone,
 averaging results from fits to the total intensity maps as well as 
the map of the core alone which was reconstructured from
the clean components.
 The errors include differences between
the maps and, 
in the case of the flux, 
differences between values measured from integration of
the core map, and values determined by  
the gaussian fit to the core.  Note that the calibration error and the pointing
error ($<$ 1 arcsecond) are not included in these quoted errors.}
              &  0.81 $\pm$ 0.03           &  1.42 $\pm$ 0.03        \\
Core Position$^c  $             &                            &                         \\
~~~~RA (h, m, s)                &  03 19 48.26 $\pm$ 0.02
    & 03 19 48.29 $\pm$ 0.03  \\
~~~~DEC ($\deg$, $\min$, $\sec$)&  41 30 45.1  $\pm$ 0.3     & 41 30 43.5 $\pm$ 0.4    \\
Core Size$^c$                   &                            &                         \\
~~~~Major Axis (arcsec)         & 4.7 $\pm$ 0.5              &   ----\footnote{
          A dash means unresolved.}    \\
~~~~Minor Axis (arcsec)         & 3.2 $\pm$ 0.5              &   ----$^c$              \\
~~~~Position Angle (degrees)    & 90 $\pm$ 10                &   ----$^c$              \\
\hline
$\alpha_{Tot}$\footnote{S$_{\nu}$ $\propto$ $\nu^{\alpha}$}
             &   \multispan2{\hfill-0.45 $\pm$ 0.28\hfill}                              \\
$\alpha_{Core}$$^e$           &     \multispan2{\hfill -0.88 $\pm$ 0.09\hfill}          \\
\hline
\end{tabular}
\end{minipage}
\end{table*}

\subsection{The AGN}
\label{agn}

The bright AGN in NGC~1275 has shown a gradual
decline of a factor 4 in brightness at 375 GHz (800 $\mum$)  between 1988 and 1992
(Stevens et~al. 1994), providing compelling evidence for the non-thermal
nature of this emission.
From our data, there is no indication for variability between
the three epochs of the present SCUBA observations to the level of
the calibration error, or about 6\%.  Moreover,
our measured 1998 total flux at 
850 $\mu$ of 1.6 Jy agrees closely with
the 800$\mum$ flux of $\sim$ 1.6 Jy 
 measured by Stevens et al. 
(from their Fig. 2) in 1994.  Thus, the
strong decline seen before 1994 does not appear to have continued over the 4
years between 1994 and 1998. 

The core is 1.8 times brighter at 850$\mum$ than at 450$\mum$, resulting in a spectral
index (Table~\ref{flux-tab})
of $\alpha$ = -0.88 $\pm$ 0.09
(S$_\nu$ $\propto$ $\nu^{\alpha}$).  This agrees with the
value found by Stevens et al. (1994) of $\alpha$ = -0.92 $\pm$ 0.17 between 375 and
150 GHz over a 6 year monitoring period. Such a steep spectral index can be
produced from
synchrotron emission.  Thus we are likely 
observing the high frequency extension
of the synchrotron emission seen at lower frequencies.  It is not possible to
extrapolate the low frequency radio flux to the sum-mm because of the variability
of the source over the different epochs of observation.  However, since
the low frequency (i.e. between $\lambda$90 cm and $\lambda$199 cm) spectral index 
within the central 1 arcminute region is dominated by a flat spectral
index core (see Pedlar et al. 1990), the total
synchrotron spectral index must turn over
between the low frequency
and sub-mm regimes.  
Given the steepness of the sub-mm spectrum,
 there is no need to postulate the presence of
a dust component in the core at any significant level.  
 A similar conclusion is reached
by Leeuw, Samson \& Robson (2000) 
for the elliptical galaxy, NGC~4374, in which sub-mm emission from
the AGN core is detected.
Note that this does not, by itself,
rule out the possibility of some dust in the core.  Indeed, the presence of
CO emission at the position of the core (Bridges \& Irwin 1998, Reuter et al. 1993)
suggests that dust is also present there.
However, the core is unlikely
to 
be dominated by dust emission or else the core spectral index would be significantly
flatter or inverted.
The sub-mm core is unresolved at 850$\mum$ but has a 
(deconvolved) size of $\sim$ 4$\arcsec$ at 450$\mum$.

If we use the core spectral index 
and extrapolate the core flux to 100$\mum$, we find a 100$\mum$
core flux of 0.22 Jy.  This is within the error bar of the total 
100$\mum$ flux of 7.9 $\pm$ 1 Jy measured by Lester et al.(1995) and
suggests that the nonthermal core 
is a negligible contributor to 
the total flux at 100$\mum$. Thus, the 100$\mum$ total flux (7.9 Jy)
can be interpreted as being entirely due to dust emission.
This agrees with Lester et al.'s
conclusion that
less than 20\% of the total 100$\mum$ emission can be due to the 
nonthermal core based on the known variability of the core and the
constancy of the 100$\mum$ flux.  Thus, the total emission changes from
being 
non-thermal core-dominated to dust-dominated between 450$\mum$ and 
100$\mum$.

\subsection{Extended Emission}
\label{extended}

The extended sub-mm emission is much
fainter than that associated with the core, but is
seen at both frequencies and appears to be real. The clean and maximum entropy maps
also show good agreement although there are some differences at low contour levels
in the 850$\mum$ maps.  
Single epoch maps were also cleaned to confirm that the
extended emission is present at all epochs to a level 
that is consistent with the shorter integration time per epoch
and higher atmospheric opacity in the July 15 observations.
The extended emission constitutes $\sim$ 33\%
of the total flux at 450$\mum$ and $\sim$ 11\% of the total flux at 850$\mum$.

The 450$\mum$ core-subtracted map 
(Fig.~\ref{final-maps-fig}c) has a much higher noise level than the 850$\mum$
map, but does show several features which remain upon smoothing to lower resolution and
which also have counterparts in the 850$\mum$ map.  The principle one is an elongated
feature to the west of the core extending northwards.  This will be referred to
as the NW extension.  Secondly, there are some
residual peaks about 45$\sec$ to the south-east of the core which again 
have a counterpart
in the 850$\mum$ map. We will refer to this region as the SE peak.
 Another peak about 1$\min$ to the south-west could possibly
be real, but this is not certain given its offset from a nearby 850$\mum$ peak.

Extended emission is most readily seen in the lower noise
850$\mum$ maps
(Fig.~\ref{final-maps-fig}d, e, and f).
These show  further 
extensions to both the east and west and ``disconnected" features elsewhere
which appear to be real, given their presence in each frame.
A fairly strong feature near the core is also seen to extend
$\sim$ 30$\sec$ to the south-west.  This latter feature also appears to be
real, though may contain a relatively larger error, given its 
proximity to the core.

The spectral index of the total emission (Table~\ref{flux-tab}) is flatter than
that of the dominant core alone, suggesting that the spectral index of 
the faint extended emission may have an inverted  (positive)
spectral index.  
 Indeed, we could measure this spectral index in 
two positions, the
NW extension and the SE peak, by first smoothing the 450$\mum$ map to
the 850$\mum$ resolution.  For the NW extension, flux measurements were made within
 a 6$\sec$ diameter region centered at
RA = 03$^h$ 19$^m$ 45$\rasec$79, DEC = 41$\deg$ 30$\min$ 57$\decsec$3, with
the resulting spectral index found to be $\alpha$ = +2.6.  
For the SE peak, the measurement was made within a 6$\sec$ diameter region centered at   
RA = 03$^h$ 19$^m$ 49$\rasec$17, DEC = 41$\deg$ 30$\min$ 00$\decsec$3,
and found to be $\alpha$ = +2.8.  Thus, where the spectral index could be measured,
the values are strongly positive.  
These are clearly dust components.

\subsubsection{Dust Temperature}

Dust is expected to be optically thin at sub-mm
wavelengths, and the temperature can, 
 in principle, be
determined via $S_\nu\,=\,\Omega\,B_\nu(T)\,Q(\nu)$ (e.g. Stevens \& Gear 2000), 
 where $S_\nu$ is the flux
density at frequency, $\nu$, $\Omega$ is the solid angle subtended by the dust,
$B_\nu(T)$ is the Planck function applicable to a dust temperature, $T$, and
$Q(\nu)$ is the frequency-dependent emissivity of the grains.  In the optically
thin limit, $Q(\nu)$ = $({{\nu}\over{\nu_0}})^\beta$, where $\beta$ is the index of
the emissivity law  and $\nu_0$ is the frequency at which
the dust becomes optically thin.  If we restrict the dust temperature
 determination for NGC~1275 to
regions for which fluxes are available at both frequencies (here, taking the
NW extension), and assume that the 
 covering factor is
the same at the two frequencies, we can estimate a temperature, adopting
a value for the unknown emissivity index, $\beta$.
 Taking $\beta$ = 1.3 as was found for M~82 
(Hughes, Gear, \& Robson 1994;
see also Leeuw et al. 2000), 
 we find a dust temperature of 19 K.  For $\beta$ = 2
(e.g. Domingue et al. 1999), we find
T = 11 K and if
 $\beta$ = 1 (Lester et al. 1995), 
T = 29 K.  Thus, for reasonable values
of $\beta$,
the dust temperature is constrained from these observations 
 to be between 10 and 30 K.

While most other flux measurements of NGC~1275 have included the
non-thermal core flux, we have noted (Sect.~\ref{agn}) 
that the core is a negligible contributor to the total 
 100$\mum$ flux.  Thus, the 100$\mum$ point should 
represent dust emission and can also be used to constrain the
dust temperature.
If we include the 100$\mum$ flux (7.9 Jy, Lester et al. 1995)
in this
analysis, however, we find that the observed flux ratios cannot
be reproduced within error bars using single values of $\Omega$,
 $\beta$ and T.  This suggests that (at least) a two component model is
required.  For $\beta$ = 1.3, the 100$\mum$/450$\mum$ ratio implies
a temperature of 38 K in comparison to 19 K found from
the 450$\mum$/850$\mum$ value.  Differences in $\beta$ or $\Omega$
are also possible.  However, we note that there is some evidence
for a two-temperature model
from a line ratio analysis of the molecular gas as well
(Bridges \& Irwin 1998).  We suspect that dust that is 
near the non-thermal core may be warmer.
 
\subsubsection{Dust Mass and Gas-to-Dust Ratio}  
\label{dust_mass}

The dust mass is determined via $M_d\,=\, S_\nu\,D^2\,/[{k_d}_\nu\,B_\nu(T)]$,
 where $D$ is the
distance to the galaxy and ${k_d}_\nu$ is the dust mass absorption coefficient.  
We will determine the dust mass using the
 850$\mum$ flux.
This is because it has typically been found (e.g. 
Alton et al. 2000) that over
90\% of observed 850$\mum$ emission arises from cold ($\ltabouteq$ 20 K) dust,
(consistent with our best temperature estimate of 19 K)
and this cold component 
constitutes the bulk of galactic dust.  Moreover,
the lower noise in the 850$\mum$ images
makes this a better choice for such a calculation.  The value of ${k_d}_\nu$ is rather
uncertain and has the same frequency dependence as $Q(\nu)$.  Again, adopting 
$\beta$ = 1.3 and extrapolating the mass absorption
coefficient from the 100$\mum$ value of
2.5 $m^2\,kg^{-1}$ (Stevens \& Gear 2000) yields ${k_d}_\nu$ = 0.15 $m^2\,kg^{-1}$.
(This is a factor of 2 higher than the value used by Dunne et al. 2000 for their
determination of dust mass functions in the local universe.)
Applying a dust temperature of 20 K
and using $S_\nu$ = 0.18 Jy (after core subtraction, see Table 2),
 we find a dust mass of
M$_d$ = 5.9$\,\times\,10^7$M$_\odot$.  
A temperature of 40 K would lower this by a factor of 2.5, using $\beta$ = 2 would
increase it by a factor of 4.5, and using the noisier 450$\mum$ 
(rather than 850$\mum$) data would lower
it by a factor of 2.6.
  Thus our best
estimate of the dust mass is 6 $\times$ 10$^7$ M$_\odot$ to within a factor of a few.
This is the first estimate of the dust mass in this galaxy from spatially resolved
emission measurements which correct for the AGN.

\begin{figure*}
\psfig{figure=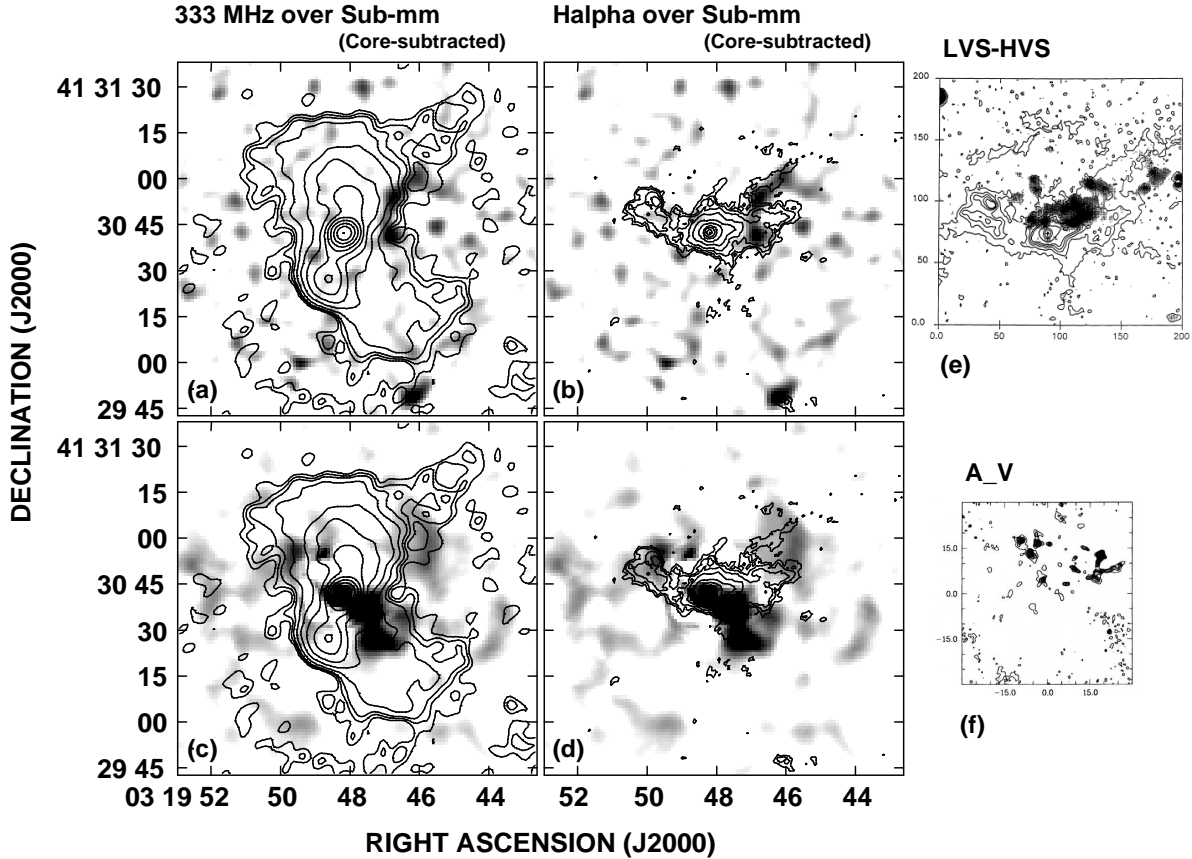,width=\textwidth}
\caption{Contours of emission at other wavebands over greyscale of the sub-mm 
emission.{\bf(a)}  333 MHz radio continuum 
(Pedlar et al. 1990) over the core-subtracted 
450$\mum$ emission, the latter smoothed to 10$\sec$ $\times$
10$\sec$ resolution. The greyscale ranges
from 27 to 90 m\jybs and contours are at 13.5, 20, 28, 50, 100, 200, 400, 800, 1600,
3000, 5000, and 8000 m\jyb.  The 333 MHz beam is 
5$\decsec$5 $\times$ 4$\decsec$9 at position angle PA = -74$\deg$.
{\bf (b)} H$\alpha$ + [NII] emission (Goudfrooij et al. 1994)
from the LVS over the same 
450$\mum$ greyscale as in (a).
Contours are in arbitrary units.
{\bf (c)}  As in (a) but the 850$\mum$ (un-smoothed) 
core-subtracted emission is the greyscale which scales from
 7 to 25m\jyb.
{\bf (d)}  Contours as in (b) with the greyscale as in (c).
{\bf (e)}  H$\alpha$ emission showing the low velocity system in contours
over the high velocity system in greyscale (Caulet et al. 1992) at the
same scale as Figs. (a) through (d).  The axes are in units of 0.406$\arcsec$.
{\bf (f)}  Optical absorption (A$_V$) from Goudfrooij et al. (1994)
at the same scale as Figs. (a) through (e).
}
\label{overlay-maps-fig}
\end{figure*}

This result can be compared with the much lower
dust mass of 7.5 $\times$ 10$^5$ M$_\odot$
(adjusted to our distance)
found for NGC~1275 by 
Goudfrooij et al. (1994) using 
optical absorption measurements. The absorbing dust
is confined to
a much smaller patchy region to the north-west of the core 
(Fig.~\ref{overlay-maps-fig}f) and
constitutes only 1.3\% of the total dust mass as determined from our 850$\mum$
measurements.  Discrepancies between the absorbing and emitting dust masses have
been noted by Goudfrooij \& de Jong (1995) who suggested that the differences
could be due to additional dust which is
distributed more broadly than in front of the stellar light distribution alone.
Our dust mass and the extended dust
distribution seen at 850$\mum$ clearly support this view and indicate that there is
much more dust in NGC~1275 seen in emission than previously determined from
absorption measurements.

Goudfrooij \& de Jong (1995) have determined dust masses for 56 Shapley-Ames 
elliptical galaxies from IRAS observations, finding dust masses in the range,  
10$^4$ to 5 $\times$ 10$^6$ M$_\odot$.  Bregman et al. (1998) have also found IRAS 
dust masses in the same range for E and S0 galaxies.  NGC~1275, therefore, has
a higher dust mass, by about an order of magnitude, than the other early-type
galaxies studied by these authors.  
However,
it is well known that IRAS observations
sample somewhat warmer dust
than SCUBA and may thereby overestimate the dust temperature, underestimating the
resulting mass (Dunne et al. 2000).  Therefore the discrepancy between the
high dust mass in NGC~1275 and the lower dust masses in E and S0 galaxies
may not be as high as an order of magnitude.  Also, these galaxies are typically
not in a cooling flow environment.
Fewer previous measurements have been made of
central cooling flow galaxies.  Edge et al. (1999), however, find masses between
3 to
5 $\times$ 10$^7$ M$_\odot$
for A1835 and A2390
(adjusting to H$_0$ = 75 km s$^{-1}$ Mpc$^{-1}$), in closer agreement
with our value for NGC~1275.

Armed with these data, we can compute a global gas/dust ratio.  For the molecular gas
mass, we use 1.6 $\times$ 10$^{10}$ M$_\odot$ (Bridges \& Irwin 1998) which uses the
Galactic 
CO to H$_2$ conversion factor, and for the atomic hydrogen mass, we use the value found
for the low velocity system by Jaffe (1990), i.e. 5 $\times$ 10$^{9}$ M$_\odot$
assuming a spin temperature of 100 K.  We find
M(HI $+$ H$_2$)/M$_d$ $\sim$ 360 with these assumptions.  This is higher
than typical Galactic values of 100 to 150 (Devereux \& Young 1990) but is consistent
with what has been found for radio quiet quasars, spiral galaxies, and
ultraluminous infra-red galaxies ($\approx$ 260 to 540; Hughes et al. 1993).

\subsection{Correlations with other Wavebands}
\label{correlations}

Figure~\ref{overlay-maps-fig}
shows overlays of the core-subtracted sub-mm maps (in greyscale, 
450$\mum$ at top and 850$\mum$ at bottom) with
emission at other wavebands.
  Note that by subtracting
the core, we cannot compare emission features within the central
$\approx$ 20$\arcsec$ diameter region.  Overlays are with the
 333 MHz radio
continuum emission 
(Fig.~\ref{overlay-maps-fig}a, c) which traces the non-thermal component,
and with the H$\alpha$ low velocity system
(Fig.~\ref{overlay-maps-fig}b, d) 
which presumably traces condensing cooling flow gas.
  For comparison, the
high velocity system, associated with the
infalling galaxy (Sect.~\ref{intro}),  is shown as greyscale with the LVS
contours in Fig.~\ref{overlay-maps-fig}e and the optical absorption
is shown in Fig.~\ref{overlay-maps-fig}f.
 
The dust emission shows some weak correlations with the 333 MHz map.
For example, the sub-mm NW extension
occurs along the outer ridge of the radio continuum emission.  There is also
weak sub-mm emission just where the radio continuum emission falls off about
1 arcminute SW of the nucleus.  A ridge of weak radio continuum
emission extending from
RA = 3$^{\rm h}$ 19$^{\rm m}$ 52$^{\rm s}$, DEC = 41$\deg$ 30$\min$ 55$\sec$ 
to 
RA = 3$^{\rm h}$ 19$^{\rm m}$ 49$^{\rm s}$, DEC = 41$\deg$ 29$\min$ 50$\sec$
is also seen in the sub-mm.  

There is no obvious correlation between the
 LVS H$\alpha$ emission and the sub-mm maps other than a possible
one between the NW extension and an H$\alpha$ extension at 450$\mum$
(Fig.~\ref{overlay-maps-fig}b).
The HVS H$\alpha$ emission (Fig.~\ref{overlay-maps-fig}e) 
occurs in a patchy distribution mainly to
the NW of and near the nucleus.  Again, there is no clear correlation
between the sub-mm emission and high velocity H$\alpha$ system
other than some correspondence in the same NW extension region.
The optical extinction map (see Fig.~\ref{overlay-maps-fig}f)
does show several knots which correspond to enhanced 450$\mum$ or
850$\mum$ emission, suggesting that this dust is in the foreground.
N{\o}rgaard-Nielsen et al. (1993) and McNamara et al. (1996)  have argued
that some of the dust, as observed in optical extinction,
corresponds to the intervening (HVS) galaxy.  

We have also investigated whether the CO emission (not shown) resembles
the sub-mm, since normally dust and molecular gas follow
similar distributions.  CO has been mapped with similar resolution 
as the sub-mm by
Bridges \& Irwin (1998) and Reuter et al. (1993); however, the most
wide-scale mapping has been out to a radius of only $\sim$ 30$\arcsec$
and little structure is visible in existing CO maps.  Bridges \& Irwin do find a
bend in the CO contours to the south-west, similar to the bright 850$\mum$
emission in that direction.  Much higher resolution CO mapping, which
cannot be directly compared to our sub-mm distribution due to the core 
subtraction
and difference in resolution, shows an east-west structure similar
to the LVS (see Inoue et al. 1996).

 Figure~\ref{colour-fig}  shows the 850$\mum$
core-subtracted map superimposed on a colour image of the ROSAT HRI X$-$ray emission
which shows a deficit of X$-$ray flux at the positions of
the radio lobes.  This morphology
has been interpreted as a displacement of the thermal
gas by the outflowing 
radio lobes (B{\"o}hringer et al. 1993, Churazov et al. 2000).  
Fig.~\ref{colour-fig}
shows the best correlation we have found with the sub-mm and
helps to explain the partial correlation 
  found between dust
and the radio continuum 
(see above) since there is a relationship between
the radio continuum  and
X$-$ray emission.  The brightest X$-$ray emission occurs in a ridge which
extends from the east of the nucleus and curves to the south.  This is
also roughly where a weak radio continuum ridge occurs (as pointed out above). 
The 850$\mum$ emission approximately follows this curved ridge 
(cf. the ``filled in" emission in the ME map, Fig.~\ref{final-maps-fig}e 
also) although the peaks
don't exactly coincide.  For example, 
dust emission curves around  the X$-$ray
peak at 
RA = 3$^{\rm h}$ 19$^{\rm m}$ 50$\rasec$5, DEC = 41$\deg$ 30$\min$ 15$\sec$.  
The bright sub-mm extension to the SW of the nucleus truncates abruptly
just where the X$-$ray emission also drops about 30$\sec$ from the nucleus;
this is 
where the SW radio lobe occurs.  To the west of the nucleus, both X$-$ray and
dust emission show an extension to $\sim$ 50 arcseconds, both ending at roughly
the same position.
 The NW extension correlates well
with an X$-$ray peak on the NW side of the galaxy with the sub-mm peak slightly
displaced to the NW of the X$-$ray peak.  Even 
some of the the very faint sub-mm emission which we at first considered to be noise
bears some resemblance to the X$-$ray.  For example, to the north of the nucleus,
the northern X$-$ray arc has 
several 850$\mum$ peaks occurring along its northern edge about 45$\sec$
from the nucleus.  About 1 arcminute from the nucleus to the NNW, 
weaker X$-$ray
emission falls off and again, weak peaks of 850$\mum$ emission can be seen
there.  

Recent observations by the Chandra satellite (Fabian et al. 2000)
confirm the general characteristics
of the X$-$ray map shown in Figure~\ref{colour-fig}.  One improvement is
in the clearer view of an X-ray absorption feature 
$\sim$ 40$\arcsec$ NW of the nucleus which is seen as a small arc just to
the west of the brightest X-ray ridge on the NW side of the nucleus (see
the low X-ray emission at RA $\sim$ 3$^{\rm h}$ 19$^{\rm m}$ 45$^{\rm s}$,
DEC $\sim$ 41$\deg$ 30$\min$ 47$\sec$).  This 
absorption feature appears to ``wrap around"
on the west side of the
brightest X-ray/sub-mm peak  NW of the nucleus and so suggests that
there may be some relation between the dust and the infalling galaxy.  However,
we do not see enhanced
sub-mm emission where the X-ray absorption occurs.  This suggests that the feature
which is producing the X-ray absorption includes some component other than
dust and/or that the characteristics of the absorbing dust are such that we are not
clearly detecting it at 450$\mum$ or 850$\mum$. 

In summary, we find that the sub-mm dust emission correlates spatially with the
X$-$ray emission, except that there may be offsets between the X-ray peaks and
sub-mm peaks (for example, 
the strongest X$-$ray peak does not have a counterpart
in the sub-mm).  


\section{Discussion}
\label{discussion}

\subsection{Source of Dust Heating}
\label{heating}

From these SCUBA data, we have estimated a dust temperature to be $\sim$ 20 K.
What are the possible heating sources for this dust?  The possibilities include
 heating 
by stellar photons, heating by hot
electrons in X$-$ray emitting gas, direct heating by X$-$rays, and heating by
the AGN.  

Since dust emission is observed to a distance of $\sim$ 1 arcminute = 20 
kpc from the AGN in NGC~1275, we can rule out AGN heating as a possible
source for this extended dust.  
Rigopoulou, Lawrence, \& Rowan-Robinson (1996) have also
looked at the submillimetre and X$-$ray data from a number of ultraluminous 
infra-red galaxies and found that most are well-fitted by a standard model of
stellar heating, even when there is evidence for an AGN.
Also, since the 450$\mum$/850$\mum$ spectral index of the
dust emission is the same for dust in the NW extension (measured
$\sim$ 6 kpc from the nucleus) as for dust at 15 kpc,
there is no need to postulate different heating
sources for the extended dust at these frequencies.  

Goudfrooij \& de Jong (1995) have considered the other heating mechanisms
for elliptical galaxies and conclude that direct heating by X$-$rays is
not feasible.  This leaves heating by stellar photons and heating by
hot electrons as the remaining viable mechanisms.
McNamara et al. (1996) have found a faint blue continuum extending out to
$\sim$ 30 kpc (1.5 arcminutes) in NGC~1275 along the filamentary H$\alpha$
system, indicating that stars do indeed exist to large distances.  
However, the close
resemblance of the sub-mm continuum in NGC~1275 to the X$-$ray structure 
and the apparent absence of a large population of stars more massive
than $\sim$ 5 $M_\odot$ (Sarazin \& O'Connell 1983,
Smith et al. 1992, McNamara et al. 1996)
 points to 
hot electrons as the source of dust heating in NGC~1275.  Lester et al.
(1995) have also shown quantitatively, that hot electrons in NGC~1275
are energetically capable of supplying the dust heating in
this galaxy.  
The fact that the 850$\mum$ emission avoids the
brightest X$-$ray peak 
suggests that dust may be more rapidly destroyed in 
this region.

\subsection{Origin of the Dust}
\label{origin}

These observations have shown that extended dust exists in NGC~1275 to very
large distances from the core ($\sim$ 20 kpc) and that the
 total dust mass is high, i.e.
6 $\times$ 10$^7$ $M_\odot$, two orders of magnitude larger than that inferred from
optical absorption studies.  
As indicated in Sect.~\ref{intro}, three possibilities for the origin of the
dust present themselves:  from
the cooling flow or in  dense clouds associated with the flow,
from ongoing star formation, or from the merger of a gas/dust rich galaxy. 
 Let us consider these possibilities.

If dust is mixed with the cooling flow gas, 
then its lifetime against destruction by sputtering is found from 
 t$_d$ =
2$\,\times\,$10$^4$ (cm$^{-3}$/$n$)($a$/0.01$\mum$), where $n$ is the density of
the hot cooling flow gas and $a$ is the grain radius (Draine \& Salpeter 1979).
The grain size is uncertain, but a reasonable upper limit is 0.1 micron 
(cf. Lester et al.
1995), and the hot gas density near the center of NGC~1275 is
0.05 - 0.1 cm$^{-3}$ (e.g. Fabian et al. 2000).
This gives t$_d$ = 2 - 4$\,\times\,$10$^6$ years. Thus, dust
will be destroyed on short timescales if it is mixed with the hot gas 
and there must be a source of dust replenishment.  The alternative
is that dust is somehow shielded from the X$-$ray environment, possibly
in dense molecular clouds for which there is (as yet) no observational
evidence.

If stellar mass loss is continuously replenishing the dust, we would 
expect this dust to be
ejected and dispersed into the X$-$ray emitting environment, suggesting that
the short lifetime determined above should be relevant.  
Determining a dust
mass loss rate from stars is not straightforward and involves a knowledge
of the initial mass function, the mass loss rates 
from stars, and the gas/dust ratio in the ejected material.
For NGC~1275,
 there currently appear to be no stars
more massive than 5 M$_\odot$ and, if more massive stars have been
produced in the past, this past massive star formation must have ceased 70 Myr
ago or more (Smith et al. 1992).  Since this is longer than the lifetime
of dust in the X-ray emitting environment, we do not expect young, massive O type
stars to contribute to the currently observed dust in NGC~1275.  
The next strongest source of mass loss, and that which will dominate in
NGC~1275, comes from 
the older, more evolved stellar population.  The remaining main sequence stars
 of 5 M$_\odot$ or less
contribute negligibly to global mass loss rates (see Snow 1982).

Both
Faber \& Gallagher (1976) and Knapp, Gunn \& Wynn-Williams (1992) have
estimated, through independent means, mass-loss rates for an evolved population
of stars.
  Knapp et al. find
$\dot{M}\,=\,2.1\,\times\,10^{-12}\,(L_{2.2\mum}/L_{\odot})\,\,M_{\odot}\,yr^{-1}$
and Faber \& Gallagher find
$\dot{M}\,=\,1.5\,\times\,10^{-11}\,(L_{B}/L_{\odot})\,\,M_{\odot}\,yr^{-1}$,
where $L_{2.2\mum}$ and
$L_{B}$ are, respectively, the
$2.2\mum$ and
 blue luminosities.  Using (from the NASA Extragalactic Database)
the total corrected blue
magnitude of 12.15, we find $\dot{M}\,=\,0.81 \,\,M_{\odot}\,yr^{-1}$
and using the ${2.2\mum}$ flux density of 159 mJy,
we find $\dot{M}\,=\,0.62 \,\,M_{\odot}\,yr^{-1}$.
The former value should be an overestimate since it includes any
contribution from the AGN and the latter should be an underestimate
since the ${2.2\mum}$ flux density is measured only within a central
32$\sec$ aperture.  Taking a mass loss rate of $0.7 M_{\odot}\,yr^{-1}$,
then, over
 2 - 4$\,\times\,$10$^6$ years, stars can contribute, at most,
$2.8\,\times\,10^6\,\,M_{\odot}$ of gas to the ISM in NGC~1275.  With
a gas/dust ratio of 360 (Sect. \ref{dust_mass}), the dust mass generated
by stars would be $8\,\times\,10^3\,\,M_{\odot}$, which is
4 orders of magnitude lower than observed. It is possible that
2 orders of magnitude could be gained by allowing dust to
survive over a timescale one order of magnitude longer, since then
there could be a contribution from hot massive stars which typically
have higher mass loss rates than the older stars. However, it is 
unlikely that 4 orders of magnitude could be gained and a dust shielding
mechanism would then also be required.
 Thus, we conclude that 
dust cannot be replenished by stellar
mass loss.  
Lester et al. (1995) came to the same conclusion for
NGC~1275 and
Goudfrooij \& de Jong (1995) have also found that 
most elliptical galaxies 
in which X$-$ray emission has
been detected contain more dust than can be accounted for by stellar mass loss.
Moreover, if the observed dust originated from stellar mass loss, 
we would expect the dust distribution to bear some resemblance to the
stellar light distribution, which is not observed.

\begin{figure*}
\psfig{figure=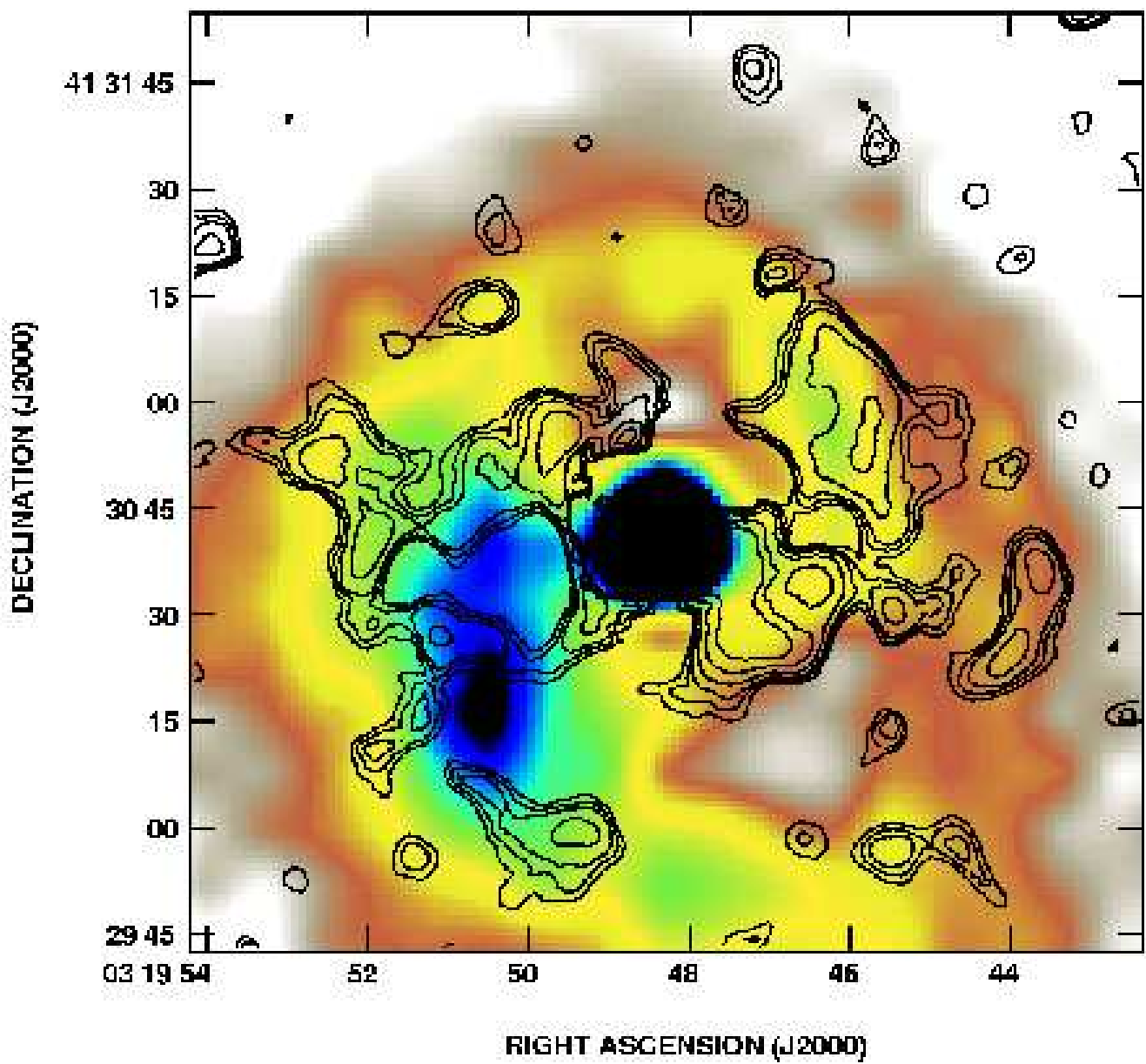,width=\textwidth}
\caption{Core-subtracted 850$\mum$ contours (as in Fig. 2f) over a colour
image of the X$-$ray emission taken from the ROSAT HRI archive
(observer, W. Voges, 1994).  The colours range
(lowest to highest intensity)
from grey through red, yellow, green, blue, violet, and black. 
 The low
intensity ``holes" north and south-west of the nucleus are filled with radio continuum
emission (see Fig. 3c); for a similar X$-$ray image, see B{\"o}hringer et al. (1993).
}
\label{colour-fig}
\end{figure*}

An obvious source of dust is the infalling galaxy.
Goudfrooij et al. (1994), in fact, make a case for 
{\it all} dust in elliptical galaxies
being supplied by external systems, and others (e.g. Pinkney et al. 1996,
Koekemoer et al. 1999)
have also supported the merger/accretion origin for dust in ellipticals.
McNamara et al. (1996) and N{\o}rgaard-Nielsen et al. (1993) 
 have suggested, for NGC~1275, that two dust systems may be present,
one associated with NGC~1275 and one associated with the the high velocity
system.  Martel et al. (1999) also find dust absorption at the 
north-west location of the HVS.
These results find some support from our observations 
since the HVS is observed to the NW of the core of NGC~1275, as is
the NW sub-mm extension.
  Thus, the 
infalling galaxy appears to contain 
dust and, one could argue, is a source of dust. 

Several lines of argument support this view.  There is accumulating
evidence that the infalling galaxy is currently interacting with NGC~1275
(Hu et
al. 1983; Unger et al. 1990; Caulet et al. 1992; N{\o}rgaard-Nielsen et al.
1993; Ferruit et al. 1998).  The HVS gas
occurs in a patchy distribution over
$\sim$ 20 kpc, in
projection (Fig.~\ref{overlay-maps-fig}e), arguing for a broad interaction
region (though the region containing dust is larger still by about a factor of 2).
The interaction timescale (impulsive
approximation) is roughly t$_i$ $\sim$ d/v, where d is the size of the target galaxy
and v is the relative velocity of the perturber. If we take the size to be
the entire projected
region over which dust emission is detected ($\sim$ 38 kpc) and
v = 3000 km s$^{-1}$, we find t$_i$ $\sim$ 1$\,\times\,$10$^7$ yrs.  This
is a factor of at least 2 higher than the dust destruction timescale, but
this may not be largely discordant, given the approximation made.
Finally, our dust mass ($6\,\times\,10^7\,\,M_\odot$) and gas-to-dust ratio
(360) are quite typical of spiral and IRAS galaxies, and
our dust temperature (20 K) is also typical of spiral galaxies though somewhat cooler than
that observed in IRAS galaxies
 (Alton et al. 2000, Dunne et al. 2000), arguing for an infalling spiral
galaxy as the origin of the dust.
In this scenario,  the dust distribution would resemble
the X-ray gas distribution (Fig.~\ref{colour-fig}) because the dust
is widely distributed and is being collisionally heated as 
 discussed
in Sect.~\ref{heating}.  The implication is that there should be additional
cold dust, as yet undetected, which is widely distributed in NGC~1275.
A difficulty with this scenario is that externally supplied dust, initially
moving at 3000 km s$^{-1}$ with respect to NGC~1275, must ``settle" into
the rest frame of NGC~1275 while the hot stars and associated 
H$\alpha$ emitting gas do not.  A theoretical development
involving an infalling spiral galaxy which also interacts with 
cooling flow gas should be brought to bear on this question. 

A final interesting idea is that the dust forms directly in the cooling flow
(Fabian et al. 1994, Hansen et al. 1995),
which would
explain the observed spatial correlation directly.  
Fabian et al. suggest that dust could form within clouds, even
in the absence of dust in the cooling flow originally, and 
would be shielded against destruction, giving it
a  longer lifetime than the above-determined value.
 The temperature
of the dust grains is expected to be is very low, with most of the dust at less
than 6 K and some as high as 11 K.  However, as the clouds become disturbed in
the inner part of the cooling flow, external heating should make the dust more
visible (and destroy the dust on short timescales again).  Since the LVS is known to
be associated with the cooling flow as well, 
however, we would also expect the dust (unless it is rapidly destroyed) to
correlate with the LVS in this scenario, for which there is no strong evidence.
 Thus, this model seems less attractive, although
should not be ruled out pending further work on disentangling density from
temperature effects in the cooling flow region.


\section{Conclusions}
\label{conclusions}

We have made SCUBA observations of the cD galaxy, NGC~1275, which is at the center
of the Perseus Cluster
cooling flow.  Emission has been detected at both 850$\mum$ and 450$\mum$ and,
  applying standard  radio astronomical techniques such as
CLEAN and Maximum Entropy to these sub-mm maps, 
we have removed structures related to
the beam shape in order to reveal real low level emission.
The resulting sub-mm distribution
can be easily separated into a bright central core and fainter
extended emission, 
allowing us to determine the spectral indices
of the core and extended emission separately.  

The core has a steep spectral index
(-0.88) suggesting 
that it can be explained primarily 
in terms of a high-frequency extension of the non-thermal AGN emission. 
The extended emission has a positive spectral index (+2.7), and is clearly
due to dust which extends as far as 20 kpc from the nucleus.  We estimate
the dust temperature to be 20 K (though 10 K to 30 K is possible, depending
on the assumed value of the dust emissivity index) and rule out the AGN 
as the main heating source,
 given the presence of dust far from the nucleus.
The dust mass is 
6 $\times$ 10$^7$ $M_\odot$ (within a factor of a few); this
dust mass is 
higher by 2 orders of magnitude than that inferred from optical extinction.
Thus there is much more dust in this galaxy than previously realized.
The gas-to-dust ratio is found to be 360.  Note that the latter value is derived
using 
an assumption of a Galactic CO to H$_2$
conversion factor as well as a 100 K spin temperature for neutral hydrogen.
Nevertheless, the derived dust mass, gas-to-dust ratio, and 
dust temperature are all typical of values being found for spiral galaxies.

A remarkable correlation between the sub-mm dust emission and the X$-$ray emission
is observed in this galaxy, although there are some
displacements in emission peaks.  When the various sources of dust heating are
considered, the strongest case can be made for collisional heating by 
hot X$-$ray emitting electrons.  This would explain the spatial correlation
between the two components.
Since the grain
destruction
timescale is short ($\sim$ 4$\,\times\,$10$^6$ yrs), continuous dust replenishment
is required to account for the sub-mm emission.  We have ruled out stars as the
source of dust and suggest instead that the infalling galaxy is the most likely
candidate, given that the observed dust parameters are typical for spiral
galaxies.
 Thus, we have a picture in which the
infalling galaxy is interacting with NGC~1275 over a large region, supplying
it with dust which is then heated most effectively by collisions with hot electrons
in the cooling flow gas.  If this scenario is correct, then more
cold dust should be present in NGC~1275 than we have detected.  
The alternative is that dust forms directly in
the cooling flow and is shielded
from destruction.
Further
spatially resolved 
multi-wavelength observations should be able to distinguish between
these scenarios by disentangling density from temperature effects.

\section*{Acknowledgments}

This research has made use of data obtained through the
High Energy Astrophysics Science Archive Research Center Online Service, provided by
the NASA/Goddard Space Flight Center.
We are grateful to 
A. Pedlar and P. Goudfrooij for providing FITS images of their
published data and also to Dr. A. Caulet for her assistance.

\bsp

\label{lastpage}

\end{document}